\documentclass{article}
\usepackage{spconf,amsmath,graphicx}
\usepackage{lipsum}

\title{The NPU-Elevoc Personalized Speech Enhancement System for ICASSP2023 DNS Challenge
}
\name{Xiaopeng Yan$^{1,2}$, Yindi Yang$^{2}$, Zhihao Guo$^{2}$, Liangliang Peng$^{2}$, Lei Xie$^{1}$*}
\address{$^1$ Audio, Speech and Langauge Processing Group, Northwestern Polytechnical University, Xi'an, China\\
$^2$ Elevoc, Shenzhen, China\\
\texttt{xiaopengyan@mail.nwpu.edu.cn, lxie@nwpu.edu.cn} \\
\texttt{\{zhihao.guo, yindi.yang, liangliang.peng\}@elevoc.com}}
\begin{document}
\ninept
\maketitle
\newcommand\blfootnote[1]{%
\begingroup
\renewcommand\thefootnote{}\footnote{#1}%
\addtocounter{footnote}{-1}%
\endgroup
}

\begin{abstract}
\vspace{-6pt}
\blfootnote{*Corresponding Author}
This paper describes our NPU-Elevoc personalized speech enhancement system (NAPSE) for the 5th Deep Noise Suppression Challenge\cite{dns2023} at ICASSP 2023. Based on the superior two-stage model TEA-PSE 2.0~\cite{ju2023tea}, our system particularly explores better strategy for speaker embedding fusion, optimizes the model training pipeline, and leverages adversarial training and multi-scale loss. According to the results\footnote{https://aka.ms/5th-dns-challenge}\footnote{https://github.com/microsoft/DNS-Challenge}, our system is tied for the 1st place in the headset track (track 1) and ranked 2nd in the speakerphone track (track 2).
\end{abstract}
\vspace{-3pt}
\begin{keywords}
personalized speech enhancement, real-time, generative adversarial network, deep learning.
\end{keywords}

\vspace{-16pt}
\section{Introduction}
\vspace{-10pt}

Taking a speaker's short enrollment as prior, personalized speech enhancement (PSE) aims to extract the target speaker's speech from a mixture signal that may contain noise, reverberation, and interfering speaker. The 5th edition of the deep noise suppression (DNS) challenge, held in ICASSP 2023, particularly focuses on PSE for full-band signal collected from headset microphone (track 1) and speaker-phone (track 2). In this challenge, our model is based on TEA-PSE 2.0~\cite{ju2023tea} -- an upgraded two-stage model from the previous DNS challenge championship~\cite{ju2022tea}. Still keeping the two-stage strategy of decomposing a difficult learning task into easier sub-processes, TEA-PSE 2.0 adopts subband operations and a time-frequency convolution module to reduce computational complexity and further improve performance. Compared to the oracle TEA-PSE 2.0, we have made substantial improvements in several aspects. First, we leverage a stronger ResNet34-based speaker embedding model which has recently achieved state-of-the-art performance on VoxCeleb~\cite{wang2022wespeaker}. Moreover, inspired by the recent observation~\cite{liu2022quantitative} that a simple filterbank feature can preserve the integrity of speaker information with good generalization, we explore the complementarity between such acoustic representation and commonly-used neural speaker embedding through fusion experiments. Second, we study the training strategy for the two-stage model with the conclusion that the best performance is achieved when we first train the stage-one model, then freeze the stage-one model to train the stage-two model, and finally, jointly train both stage models. Ultimately, we improve model optimization with multi-scale loss~\cite{kim2021se} and GAN loss~\cite{fu2021metricgan+,fu2022metricgan}. Particularly, we adopt MetricGAN~\cite{fu2021metricgan+,fu2022metricgan} to predict PESQ and DNSMOS because it can learn these non-differentiable metrics to optimize the model.

\vspace{-12pt}
\section{Proposed Method}

\vspace{-12pt}
\subsection{Overview}
\vspace{-6pt}
Our system is mainly composed of three parts: speaker encoder, speech enhancement model, and MetricGAN discriminator. A RestNet34-based speaker encoder~\cite{wang2022wespeaker} extracts speaker embedding from the enrollment speech. The speaker embedding, together with utterance-level mean and standard deviation of the Fbank feature extracted from the enrollment speech, goes through a learnable fusion module. The fused embedding is thus fed into the speech enhancement model as the prior information for speaker extraction.

The speech enhancement model follows that in TEA-PSE 2.0~\cite{ju2023tea}, consisting of a two-stage model and subband operations. The two-stage model includes MAG-Net (stage one) and COM-Net (stage two) to process magnitude and complex features respectively. The encoder of MAG-Net is composed of three frequency down-sampling (FD) layers, the decoder of MAG-Net is composed of three frequency up-sampling (FU) layers, and the middle layer is composed of four stacked gated temporal convolutional modules (S-GTCM). Likewise, COM-Net has a similar network topology as MAGNet, while its dual-decoder architecture is designed to estimate the real and imaginary spectrum separately. The MetricGAN discriminator is introduced to help the speech enhancement model to better optimize perceptual metrics including PESQ and DNSMOS.

\vspace{-15pt}
\begin{figure}[t]
    \centering
    \includegraphics[scale=0.60]{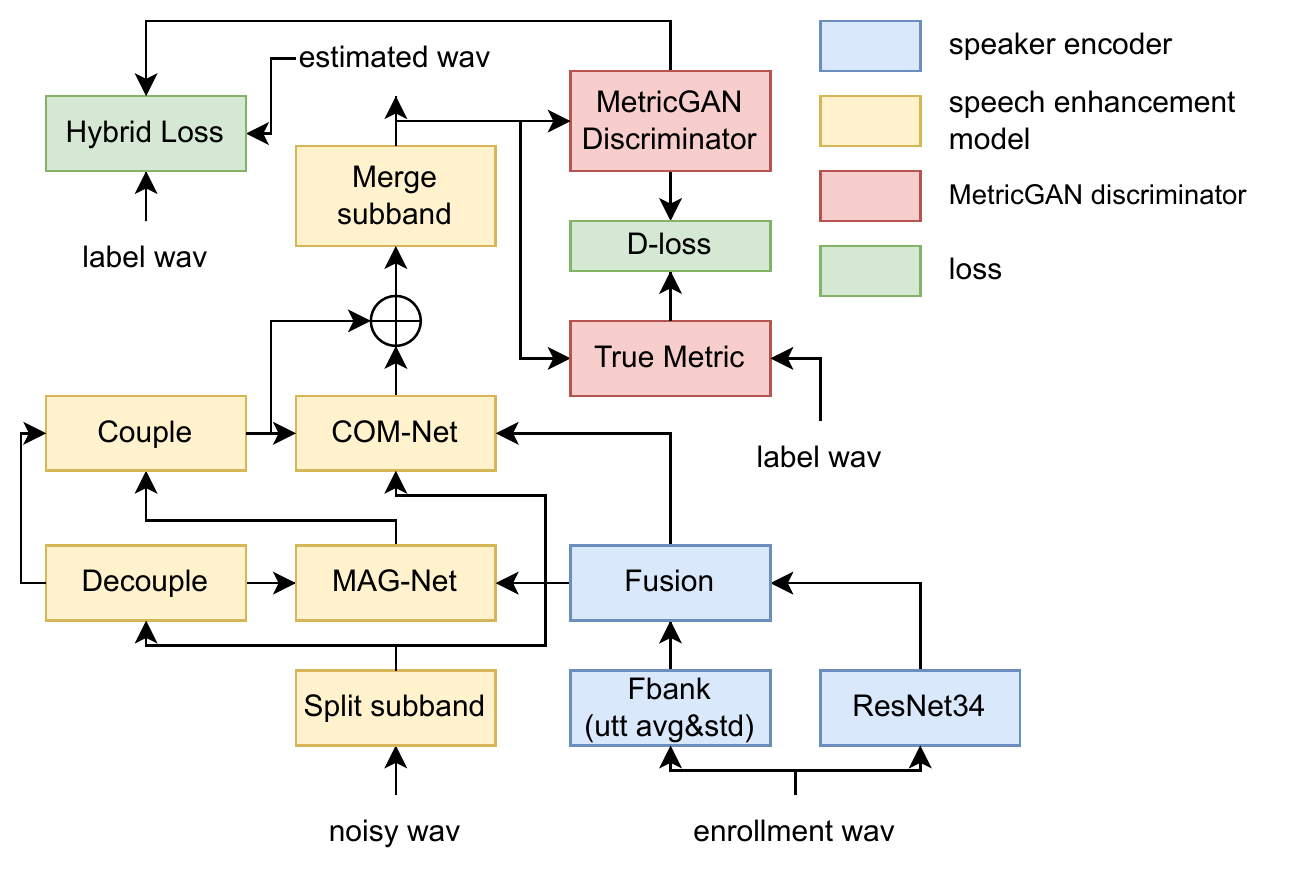}
    \vspace{-15pt}
    \caption{Architecture of the NAPSE system.}
    \vspace{-20pt}
\end{figure}

\vspace{-6pt}
\subsection{GAN}
\vspace{-8pt}
Adversarial learning has shown superior performance recently. Perceptual metrics that are closer to human auditory impression on speech quality, have been recently considered as optimization targets~\cite{fu2022metricgan,bai2022perceptual}. Thus, we adopt GAN to learn these non-differentiable metrics. Specifically, we use MetricGAN+~\cite{fu2021metricgan+} to predict PESQ and MetricGAN-U~\cite{fu2022metricgan} to predict DNSMOS OVRL respectively. 


\vspace{-12pt}
\begin{equation}
\begin{aligned}
m_{\text{PESQ}}(\hat{s},s) &= (\text{PESQ}(\hat{s},s)+0.5)/5
\\
m_{\text{DNSMOS}}(\hat{s},s) &= (\text{DNSMOS}(\hat{s})-1.0)/4
\\
\mathcal{L}_{D} &= \left| D(\hat{s}) - m(\hat{s},s) \right|^2
\\
\mathcal{L}_{G} &= \left| D(\hat{s}) - 1.0 \right|^2
\end{aligned}
\end{equation}

where $s$ and $\hat{s}$ refer to the original clean and estimated sources. $m(\hat{s},s)$ is the function to get metric, and values are normalized to $[0,1]$. $\mathcal{L}_{D}$ is used to train the MetricGAN discriminator network and $\mathcal{L}_{G}$ is used to train the speech enhancement network. 

\vspace{-12pt}
\subsection{Loss function}
\vspace{-6pt}





For the stage-one network, we use the scale-invariant signal-to-noise ratio (SI-SNR) loss $\mathcal{L}_\text{si-snr}$, the asymmetric loss $\mathcal{L}_\text{asym}$ and the magnitude spectral mean square loss $\mathcal{L}_\text{mag}$. For the stage-two network, we add the complex spectral mean square loss $\mathcal{L}_{\text{RI}}$. In addition to the above losses in TEA-PSE 2.0, we add multi-scale loss~\cite{kim2021se} and GAN loss as well, which are formulated as



\vspace{-9pt}
\begin{equation}
\begin{aligned}
\mathcal{L}_{1}&=\mathcal{L}_\text{si-snr} + \frac{1}{M}\sum_{m=1}^{M}({\mathcal{L}_\text{asym} + \mathcal{L}_\text{mag}}) + \mathcal{L}_{G}
\\
\mathcal{L}_{2}&=\mathcal{L}_{\text{si-snr}} + \frac{1}{M}\sum_{m=1}^{M}({\mathcal{L}_{\text{asym}} + \mathcal{L}_{\text{mag}} + \mathcal{L}_{\text{RI}}}) + \mathcal{L}_{G}
\end{aligned}
\end{equation}

\vspace{-4pt}
where $\mathcal{L}_{1}$ and $\mathcal{L}_{2}$ are the total loss of stage one and two models respectively, and $m$ indicates the scale corresponding to different STFT configurations. In this paper, we set $M=3$ with number of FFT bins $\in$ \{512, 1024, 2048\}, hop sizes $\in$ \{240, 480, 960\}, and window lengths $\in$ \{480, 960, 1920\} respectively.

\vspace{-12pt}
\section{Experiments}

\vspace{-12pt}
\subsection{Dataset}
\vspace{-6pt}
Speech and noise data are all taken from DNS4 track2 personalized dataset and DNS5 dataset. We simulate 150,000 RIR clips by image method\footnote{https://github.com/phecda-xu/RIR-Generator} and the RT60 of RIRs ranges from 0.1s to 1.2s. Totally we simulate about 2,000 hours of data for the model training of the two tracks while 170 hours are used as a subset for quick ablation study on \textit{dev\_test} set. The model structures for the two tracks are the same but trained using different sets on the official Github.

\vspace{-12pt}
\subsection{Training Setup}
\vspace{-6pt}
During the model training, the parameters of Resnet34 speaker encoder are always frozen. The learnable fusion module is a 256-dimensional Dense layer to combine 256-dimensional Resnet34 embedding and 160-dimensional Fbank (80 means and 80 standard deviations). The configuration of the speech enhancement model is exactly the same as that of TEA-PSE 2.0~\cite{ju2023tea}. The subband split and merge is based on PQMF and the number of subbands is set to 4. Window length and hop size are 20ms and 10ms respectively. The final submitted model has 12.49M parameters in total, where the parameters of the Resnet34 part is 6.63M. The parameters of the MetricGAN discriminator are not included here as it is discarded during inference. The multiply-accumulate operations (MACs) per second are 8.5G. The average real-time factor (RTF) per frame is 0.48, tested on Intel(R) Xeon(R) Gold 5218 CPU @ 2.30GHz using a single thread for ONNX inference testing.

\vspace{-12pt}
\subsection{Results and analysis}
\vspace{-6pt}
We first perform a quick ablation study based on the models trained using the 170-hour subset. Stage training test results on the oracle TEA-PSE2.0 model is shown in Table~\ref{tab:figure2}. We can see that the best performance is achieved when we first train the stage-one model, then freeze the stage-one model to train the stage-two model, and finally, jointly train the both stage models. We then discard the stage-one model and use the stage-two model only for a quick investigation, as shown in Table~\ref{tab:figure1}. We can see that the use of Fbank for speaker embedding fusion, multi-loss, and perceptual metric-based adversarial learning are also beneficial to the performance. Specifically, for the use of perceptual metrics, using MetricGAN-U to estimate DNSMOS SIG and BAK is better than estimating DNSMOS OVRL and PESQ. Here DNSMOS SIG and BAK are estimated each using a separate MetricGAN-U. Table~\ref{tab:figure3} shows the final subjective listening results of the challenge. Our system is tied for 1st place in Track 1 and ranked 2nd in Track 2 according to the final score.



\begin{table}[!t]
    \centering
    \caption{Training strategy on the \textit{dev\_test} set.(P.835 pDNSMOS)}
        \footnotesize
    \label{tab:figure2}
    \begin{tabular}{lcccccc}
    \hline
        ~ & \multicolumn{3}{c}{Track 1 headset} & \multicolumn{3}{c}{Track 2 non-headset} \\ 
        ~ & SIG & BAK & OVRL & SIG & BAK & OVRL \\ \hline
        Noisy & \textbf{3.98} & 2.48 & 2.68 & \textbf{4.14} & 2.34 & 2.68 \\
        Stage-1 & 3.75 & 3.82 & 3.24 & 3.85 & 3.95 & 3.41 \\ 
        ~~Stage-2 & 3.72 & 3.94 & 3.28 & 3.80 & 4.09 & 3.43 \\ 
        ~~~~Joint 1\&2 & 3.81 & \textbf{3.99} & \textbf{3.38} & 3.92 & \textbf{4.17} & \textbf{3.56} \\ 
    \hline
    \end{tabular}
    \vspace{-18pt}
\end{table}

\begin{table}[!t]
    \centering
    \caption{Ablation study on the \textit{dev\_test} set. (P.835 pDNSMOS)}
        \footnotesize
    \label{tab:figure1}
    \begin{tabular}{lcccccc}
    \hline
        ~ & \multicolumn{3}{c}{Track 1 headset} & \multicolumn{3}{c}{Track 2 non-headset} \\ 
        ~ & SIG & BAK & OVRL & SIG & BAK & OVRL \\ \hline
        Noisy & 3.98 & 2.48 & 2.68 & 4.14 & 2.34 & 2.68 \\
        Stage-2 only & 3.84 & 3.65 & 3.23 & 3.97 & 3.74 & 3.40 \\
        ~~+Fbank & 3.87 & 3.71 & 3.27 & 3.98 & 3.84 & 3.45 \\
        ~~+Multi-loss & 3.84 & 3.73 & 3.27 & 3.97 & 3.82 & 3.43 \\
        ~~+PESQ & 3.83 & 3.76 & 3.28 & 3.98 & 3.88 & 3.47 \\
        ~~+OVRL & 3.85 & 3.72 & 3.27 & 3.99 & 3.81 & 3.45 \\
        ~~+SIG\&BAK & 3.88 & 3.74 & 3.30 & 4.01 & 3.89 & 3.50 \\
    \hline
    \end{tabular}
    \vspace{-18pt}
\end{table}

\begin{table}[!t]
    \centering
    \caption{DNS5 challenge official results for our system.}
        \footnotesize
    \label{tab:figure3}
    \begin{tabular}{lccccc}
    \hline
        \multicolumn{6}{c}{Track 1 headset}  \\
        ~ & SIG & BAK & OVRL & WACC & SCORE \\ \hline
        Noisy & \textbf{3.76} & 1.22 & 1.22 & \textbf{0.843} & 0.449 \\
        Baseline\_p & 3.20 & 2.67 & 2.34 & 0.687 & 0.511 \\
        NAPSE & 3.58 & \textbf{2.87} & \textbf{2.69} & 0.758 & \textbf{0.590} \\
    \hline
        \multicolumn{6}{c}{Track 2 non-headset} \\
        ~ & SIG & BAK & OVRL & WACC & SCORE \\ \hline
        Noisy & \textbf{3.83} & 1.22 & 1.24 & \textbf{0.857} & 0.459 \\
        Baseline\_p & 3.22 & 2.68 & 2.38 & 0.727 & 0.537 \\
        NAPSE & 3.60 & \textbf{2.78} & \textbf{2.58} & 0.769 & \textbf{0.581} \\
    \hline
    \end{tabular}
    \vspace{-16pt}
\end{table}




\footnotesize
\bibliographystyle{IEEEbib}
\vspace{-9pt}
\bibliography{refs}

\end{document}